\documentclass[12pt]{iopart}

\usepackage{graphicx,amsfonts,color,lineno}

\newcommand{\ket}[1]{\ensuremath\,|{#1}\rangle}
\newcommand{\bra}[1]{\ensuremath\langle #1 |\,}

\renewcommand{\rho}{\varrho}

\begin{document}

\title[A macrorealistic test in hybrid quantum optomechanics]{A macrorealistic test in hybrid quantum optomechanics}

\author{Marta Maria Marchese$^1$, Hannah McAleese$^1$, Angelo Bassi$^{2,3}$, 
Mauro Paternostro$^1$}

\address{$^1$ School of Mathematics and Physics, Queen's University, Belfast BT7 1NN, United Kingdom}
\address{$^2$ Department of Physics, University of Trieste, Strada Costiera 11 34151, Trieste, Italy}
\address{$^3$ Istituto Nazionale di Fisica Nucleare, Trieste Section, Via Valerio 2 34127, Trieste, Italy}

\begin{abstract}

We discuss a scheme for macrorealistic theories of the Leggett-Garg form [A. J. Leggett and A. Garg, Phys. Rev. Lett. {\bf 54}, 857 (1985)]. Our scheme is based on a hybrid optomechanical system.
It seems reasonable to test these inequalities with an optomechanical system, since in an optomechanical cavity it is possible to create non-classical states of the mirror through a projective measurement on the cavity field.
We will present the protocol to generate such non-classicality for a general optomechanical cavity and after we will carry out a theoretical test for one of the possible formulations of these inequalities using a hybrid  optomechanical system. Specifically, the inequality will be investigated for an harmonic oscillator coupled to a two-level system, which replaces the light field of the cavity. The aim is to reproduce, with this system, the evolution of a single spin-1/2 for which the inequality is violated; this is achievable through the conditioning of the two-level system which will be used as an ancilla.
\end{abstract}

%
\vspace{2pc}
\noindent{\it Keywords}: Leggett-Garg, optomechanics, foundations, temporal correlations

%
\submitto{\jpb}
%
\maketitle
%
%

\section{\label{sec:level1}Introduction}

Contrarily to what one might naively believe, the observation of distinct quantum superpositions is an insufficient step to exclude a realistic picture at the macroscopic level. Temporal Bell-like inequalities such as those originally proposed by Leggett and Garg~\cite{LG} provide {\it the} avenue to ascertain the presence of macroscopic quantum coherence in the state of a given system. 

Such inequalities make use of the framework built around two assumptions that, together, define classical (macroscopic) reality. The first claims that  measurements can be made on a system without affecting its subsequent evolution. This embodies the ``non-invasive measurability" assumption. The second states that, at any instant of time, the system itself will be in a well-defined state among those it has available, thus providing the ``macroscopic realism" assumption. The simultaneous validity of such assumptions constrains very strongly the values that the two-time auto-correlation function of suitably chosen observables of the system can take. 

The values taken at different times by such function can be organized in the form of inequalities~\cite{LG}, akin to Bell's ones, providing benchmarks for any dynamics conforming to our classical intuition. The violation of such inequalities rules out the framework defined by the two assumptions above and that is commonly intended as {\it macrorealism}. The experimental falsification of macrorealistic inequalities has been recently reported in setups based on linear optics~\cite{Goggin2011, Dressel2011}, nuclear magnetic resonance~\cite{Souza2011,Athalye2011,Souza2013}, superconducting quantum circuits~\cite{Palacios2010}, spin impurities in silicon \cite{Knee2012}, a nitrogen-vacancy defect in diamond~\cite{Waldherr2011}, and a single atom loaded in an optical lattice and subjected to quantum-walk dynamics~\cite{Robens2015,Robens2016}. Macrorealistic arguments have been used in reference~\cite{Wilde2010} to investigate non-classicality of excitation-transfer processes in light-harvesting complexes.
Experimental tests of the LG inequalities, going beyond two-level systems, have been implemented with continuous modular variable measurements of the motional degree of freedom of a trapped ion~\cite{fluhmann}.

However, to date, the falsification of macrorealistic inequalities has addressed nearly exclusively microscopic systems.
A test fit to address non-classical features in the dynamical evolution of genuinely mesoscopic/macroscopic systems is not only much needed but also fundamentally interesting: the macroscopicity of a system increases with the mass of the system itself and with the degree of distinguishability of the components of a superposition state, which might require, for instance, large spatial separations achievable only in truly large-scale systems. Only few recent efforts have moved in this direction, using dichotomic measurement of oscillators \cite{bose}.

A promising avenue towards the narrowing of the experimental falsification of macrorealistic theories is provided by quantum optomechanics, where the motion of massive mechanical systems is driven, controlled and detected by their suitably arranged coupling to optical modes of a cavity~\cite{Aspelmeyer}. Schemes for the preparation of large-size superposition states of the mechanical system have been drawn~\cite{Pikovski,Brunelli2018}. Progress towards the control of hybrid systems embedding effective two-level systems into otherwise standard optomechanical platforms offer additional leverage potential for the engineering of non-classical states of massive mechanical systems~\cite{Vacanti2013,Rabl2009,Armour2002,Tian2005,Arcizet2011,Kolkowitz2012}.

This paper addresses precisely this point by proposing a scheme that, by making use of controlled dynamics in a hybrid optomechanical system and a special information encoding protocol inspired by coherent state-based quantum computing~\cite{Jeong2002}, is able to falsify macrorealistic inequalities of the Leggett-Garg form.

The remainder of this manuscript is organized as follows: in section~\ref{sec:level2} we review the arguments behind the construction of a Leggett-Garg-like inequality for a single spin-1/2 particle. This will provide the benchmark system for the analysis that we present in later sections. section~\ref{sec:level3} introduces the hybrid optomechanical platform that we use to test macrorealism in the dynamics of a genuinely massive mechanical system. We discuss both the effective dynamical map operated on the mechanical system and the special encoding of information that we propose to make our scheme akin to the spin-1/2 case of section~\ref{sec:level2}. The falsification of a macrorealistic inequality is then discussed in section~\ref{sec:level4}. Finally, in section~\ref{sec:level5} we draw our conclusions. 

\section{\label{sec:level2}Leggett-Garg inequalities}

Here, we briefly discuss the form of the macrorealistic inequality addressed in our study and provide an explicit example based on the dynamics of a simple spin-1/2 particle. 

Starting from the assumptions of non-invasive measurability and macrorealism per se, Leggett and Garg predicted that the two-time autocorrelation functions for a 
dichotomic observable $\hat{Q}$ (whose only measurement outcomes are $Q(t_{i})\in\lbrace+1,-1\rbrace$) of any physical system are constrained to satisfy the Leggett-Garg (LG) inequality~\cite{LG}
\begin{equation}
\label{LGI}
K\equiv|C_{01}+C_{12}+C_{23}-C_{03}|\leq2.
\end{equation}
Here, $C_{ij}$ is the 
two-time correlation function between two measurements at discrete times $t_{i}$ and $t_{j}$, defined as
\begin{equation}
C_{ij}=\sum_{Q_{i}=\pm1}\sum_{Q_{j}=\pm1}Q_{i}Q_{j} P_{ij}^{Q_{i},Q_{j}}.
\end{equation}
Here, $P_{ij}^{Q_{i},Q_{j}}$ is the joint probability of obtaining the outcomes $Q_{i}{=}\pm1$ ($Q_{j}{=}\pm1$) at the time $t_{i}$ ($t_{j}$). Equation~(\ref{LGI}) thus entails a sequence of measurements performed at times $t_i~(i=1,..,3)$. In fact, the measurement at the initial time $t_0$ can be absorbed in the process of initial-state preparation, and thus bypassed. Any classical (and thus macrorealistic) dynamics results in a function $K$ that satisfies equation~(\ref{LGI}) whose violation signals the departure from the framework set by the assumptions underpinning macrorealism. Suitable quantum dynamics violates the macrorealistic constraint, as will be explicitly shown in section~\ref{sec:level2a}. A simple derivation of the constraint set by equation~(\ref{LGI}) is given in reference~\cite{Emary}, while in reference~\cite{Le2017} the links between divisibility of the underlying dynamics and the violation of temporal Bell-like inequalities akin to equation~(\ref{LGI}) have been investigated.   

\subsection{Example: Falsification of macrorealistic theories using a two-level system}
\label{sec:level2a}
Let us now provide a benchmark example of quantum dynamics that violates the macrorealistic boundary. For the purpose of this work, it is enough to consider a simple two-level system (such as a spin-1/2 particle) evolving according to the Hamiltonian $\hat{H}=\omega\hat{\sigma}_{x}$ and probed by the dichotomic observable $\hat{Q}=\hat{\sigma}_{z}$. Here $\hat\sigma_{j}$~($j=x,z$) is the $j$-Pauli matrix of the two-level system and $\omega$ is the frequency splitting between the logical states $\ket{\pm1}$ such that $\hat Q\ket{\pm1}=\pm\ket{\pm1}$. We assume the initial state of the system to be $\ket{+1}$, although any other initial state would result in the same conclusions as the following. 

The two-time autocorrelation function $C_{ij}$ reads explicitly
\begin{eqnarray}
C_{ij}&=P^{++}_{ij}+P^{--}_{ij}-P^{+-}_{ij}-P^{-+}_{ij}\nonumber\\
&=P^+_j(P^{+|+}_{ij}-P^{-|+}_{ij})+P^-_j(P^{-|-}_{ij}-P^{+|-}_{ij})\\
&=P^+_j(P^{+|+}_{ij}-P^{-|+}_{ij})+(1-P^+_j)(P^{-|-}_{ij}-P^{+|-}_{ij})\nonumber,
\end{eqnarray}
Here we have streamlined the notation by setting $P_{ij}^{Q_{i},Q_{j}}\equiv P_{ij}^{\alpha\beta}$ with $\alpha,\beta=\pm1$. Moreover, $P^\pm_i$ is the probability to have outcome $\pm1$ at time $t_i$, while $P^{\alpha|\beta}_{ij}$ is the conditional probability to get outcome $\alpha$ at time $t_i$ provided that outcome $\beta$ was obtained at measurement time $t_j$. We have used the completeness relation of probability $P^+_i+P^-_i=1$.

\begin{figure}
    \centering
    \includegraphics[width=0.65\columnwidth]{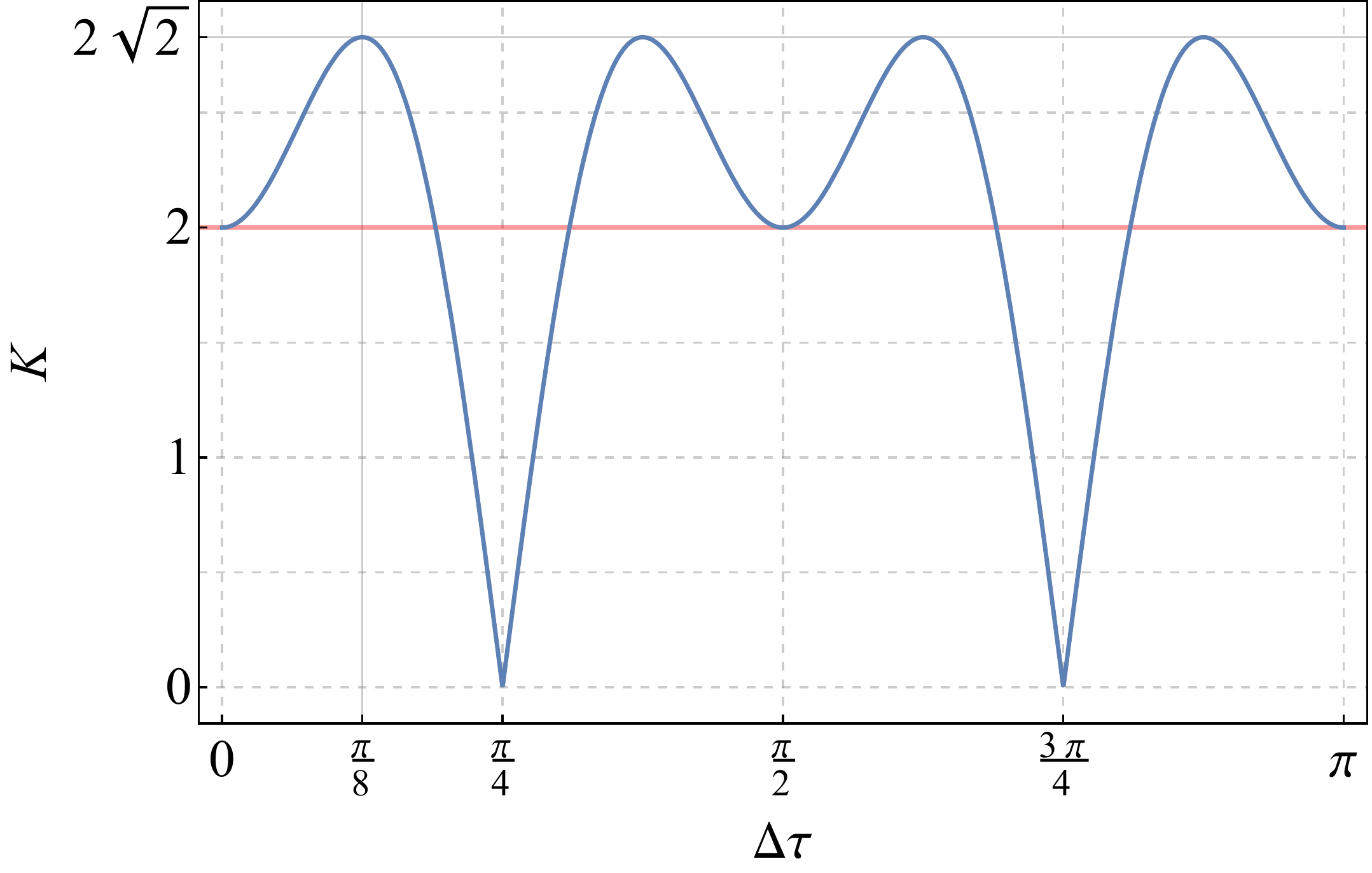}
    \caption{Leggett-Garg function $K$ against the time interval $\Delta\tau$. The function is periodic in time with period ${\pi}/{2}$, and the macrorealistic bound of 2 is  violated in the time interval $0<\Delta\tau<0.59831$. Maximum violation ($K=2\sqrt{2}$) is achieved for $\Delta\tau={\pi}/{8}$.}
    \label{spinplot}
\end{figure}

While, in principle, the instants of time at which the measurements should be performed are entirely arbitrary, it is convenient to take equally spaced values of $t_j~(j=0,..,3)$ and call $\omega(t_{j+1}-t_j)=\Delta\tau$. Under such conditions, we have $C_{01}=C_{12}=C_{23}=\cos(2\Delta\tau)$ and $C_{03}=\cos(6\Delta\tau)$, so that 
\begin{equation}
\label{LGspin}
K(\Delta\tau)=|3 \cos(2\Delta\tau)-\cos(6\Delta\tau)|.
\end{equation}
The plot of $K(\Delta\tau)$ against the -- so far undetermined --  value of $\Delta\tau$ is given in figure~\ref{spinplot}, which reveals the existence of time windows within which $K(\Delta\tau)\in[2,2\sqrt2]$. The shortest time at which the maximum violation of the macrorealistic bound is achieved is $\Delta\tau=\pi/8$. This is remarkable as, for this value, the state resulting from the evolution $e^{-i\Delta\tau \hat\sigma_x }=\cos(\Delta\tau)I-i\sin(\Delta\tau)\hat\sigma_x$ is $0.924\ket{+1}-i0.383\ket{-1}$, i.e. a state with quantum coherence -- defined in terms of the $l_1$-norm of coherence~\cite{baumgratz} -- as small as $0.708$. 

We will use this example as a benchmark for the protocol based on optomechanical dynamics illustrated in section~\ref{sec:level3}.

\section{Dynamics and control of a hybrid optomechanical system}
\label{sec:level3}

In this section we will first illustrate the hybrid optomechanical setting that we will use for the design of the test of macroscopic realism at the core of this work. We will then draw a comparison between the dynamics resulting from the form of control that we propose and what has been described in section~\ref{sec:level2a}.

\subsection{A hybrid optomechanical system}
\label{sec:level3a}

\begin{figure}
\centering
\hskip1.cm{\bf (a)}\hskip5.5cm{\bf (b)}\\
    \includegraphics[width=0.65\columnwidth]{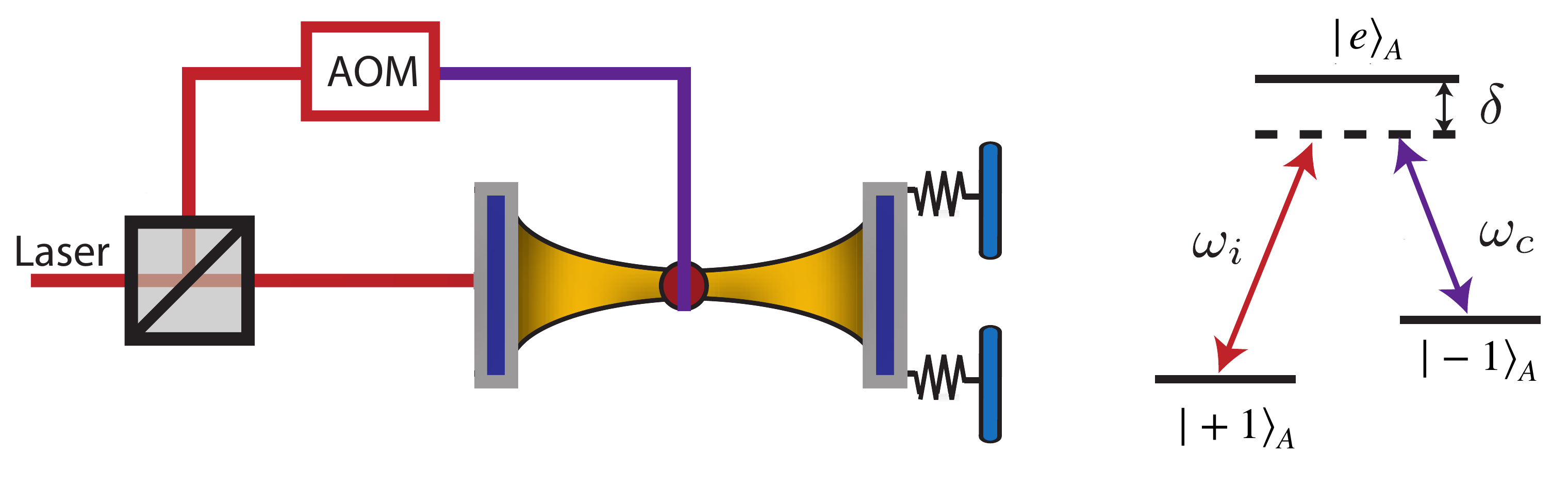}
    \caption{Panel {\bf (a)} shows the scheme of principle for the engineering of the effective interaction Hamiltonian in equation~(\ref{hamiltonian}): An optomechanical cavity embeds a three-level atom whose energy scheme is shown in panel {\bf (b)}. The parameters entered in the figure are introduced in the body of the manuscript. The energy difference between states $\ket{+1}_A$ and $\ket{-1}_A$ is $\omega_{-+}$. We show the symbol used for an acousto-optic modulator (AOM) that can be used to generate the radial field (driving one of the atomic transitions) directly from the pump that drives the optomechanical cavity. }
    \label{system}
\end{figure}

Although various controlled-coupling schemes can be considered, depending on the specific platform that one aims to manage, the system that we consider here is based on the tripartite coupling between a three-level atom in a $\Lambda$ configuration, a single-mode optical cavity pumped by a laser field at frequency $\omega_p$ and the movable mirror of an optomechanical cavity~\cite{Vacanti2013}. The atom is driven by an external field at frequency $\omega_i$ that enters the cavity radially, as in figure~\ref{system}. We label $\{\ket{+1}_A,\ket{-1}_A\}$ the states belonging to the fundamental atomic doublet and $\ket{e}_A$ the excited state. The atomic transition $\ket{+1}_A\leftrightarrow\ket{e}_A$ ($\ket{-1}_A\leftrightarrow\ket{e}_A$) is guided, at rate $\Omega$ ($g$), by the radial field (the cavity field). The detuning between each transition and the respective driving field is $\delta$, while $\Delta=\omega_c-\omega_p$ is the cavity-pump detuning. The movable mirror is schematized as a harmonic oscillator with frequency $\omega_m$, coupled to the cavity field through radiation-pressure~\cite{Aspelmeyer}. We assume the conditions $\delta\gg\Omega\gg{g}\gg\gamma_e$ with $\gamma_e$ the decay rate from the atomic excited state, such that an off-resonant two-photon Raman transition is realized. In the rotating frame defined by the operator $\omega_p\hat a^\dag\hat a + \omega_i \ket{e}\bra{e}_A + \omega_{-+}\ket{-1}\bra{-1}_A$, where $\hat a$ ($\hat a^\dag$) is the annihilation (creation) operator of the cavity field and $\hat b$ ($\hat b^\dag$) is the corresponding operator of the mirror, the Hamiltonian of the system reads  
$\hat H_{sys} = \hat H_A + \hat H_{R} + \hat H_{M} + \hat H_{C} + \hat H_{MC} + \hat H_{CP}$, where
\begin{eqnarray}
&\hat H_A=\delta \ket{e}\bra{e}_A,\quad\hat H_M=\omega_m\hat b^\dag\hat b,\quad\hat H_C=\Delta\hat a^\dag\hat a,\\
&\hat H_{MC}=\chi\hat a^\dag\hat a(\hat b^\dag+\hat b),\\
&\hat H_R=\Omega\ket{e}\bra{+1}_A+ge^{i\delta t}\hat a^\dag\ket{-1}\bra{e}_A+h.c.
\end{eqnarray}
Here, $\hat H_A$ is the atomic part of the energy, $\hat H_R$ is the Raman Hamiltonian, $\hat H_M$ ($\hat H_C$) is the free Hamiltonian of the mirror (cavity field) and $\hat H_{MC}$ is the radiation-pressure term. Finally, $\hat H_{CP}$ is the cavity-pump interaction. We also take $\Delta\gg g,\chi$, so that both the atomic excited state and the cavity field are only virtually populated and can be eliminated from the dynamics of the system. This leads to the effective Hamiltonian
\begin{equation}
\label{hamiltonian}
\hat H_{eff}=\eta\ket{{+1}}\bra{+1}_A(\hat b^\dag+\hat b)_M.
\end{equation}
with $\eta = \chi g^2 \Omega^2/\delta^2\Delta^2$. This derivation can be found in reference~\cite{Vacanti2013}. Through the two-photon Raman transition, the virtual quanta resulting from the atom-cavity field interaction are transferred (by the bus embodied by the cavity field) to the mechanical system. Therefore, the state of the latter experiences a displacement (in phase space) conditioned on the state of the effective two-level atomic system resulting from the elimination of the excited state. Remarkably, this mechanism allows for the independent preparation of the atomic and mechanical subsystems: the mechanism in equation~(\ref{hamiltonian}) can be turned off by a suitably large two-photon Raman detuning $\delta$. 

\subsection{Time evolution map}
\label{sec:level3b}

Having introduced the effective model that we aim at exploiting, we now describe the controlled map that will be used to mimic the dynamics of the spin-1/2 particle illustrated in section~\ref{sec:level2a}. 

We assume the initial state of the atomic two-level system to be an arbitrary superposition $\ket{\psi_0}_A=c_{-}\ket{-1}_A+c_{+}\ket{+1}_A$ of the energy states $\ket{\pm1}_A$ ($|c_{-}|^2+|c_{+}|^2=1$), while the analogous state of the mechanical system is the coherent state $\ket{\alpha}_M$ (with $\alpha\in\mathbb{C}$ being the amplitude of the coherent state). Such states can be engineered with large purity using pulsed-driving schemes as in references~\cite{Pikovski,Rossi}.

\begin{figure}[t]
\centering
    \includegraphics[width=0.5\columnwidth]{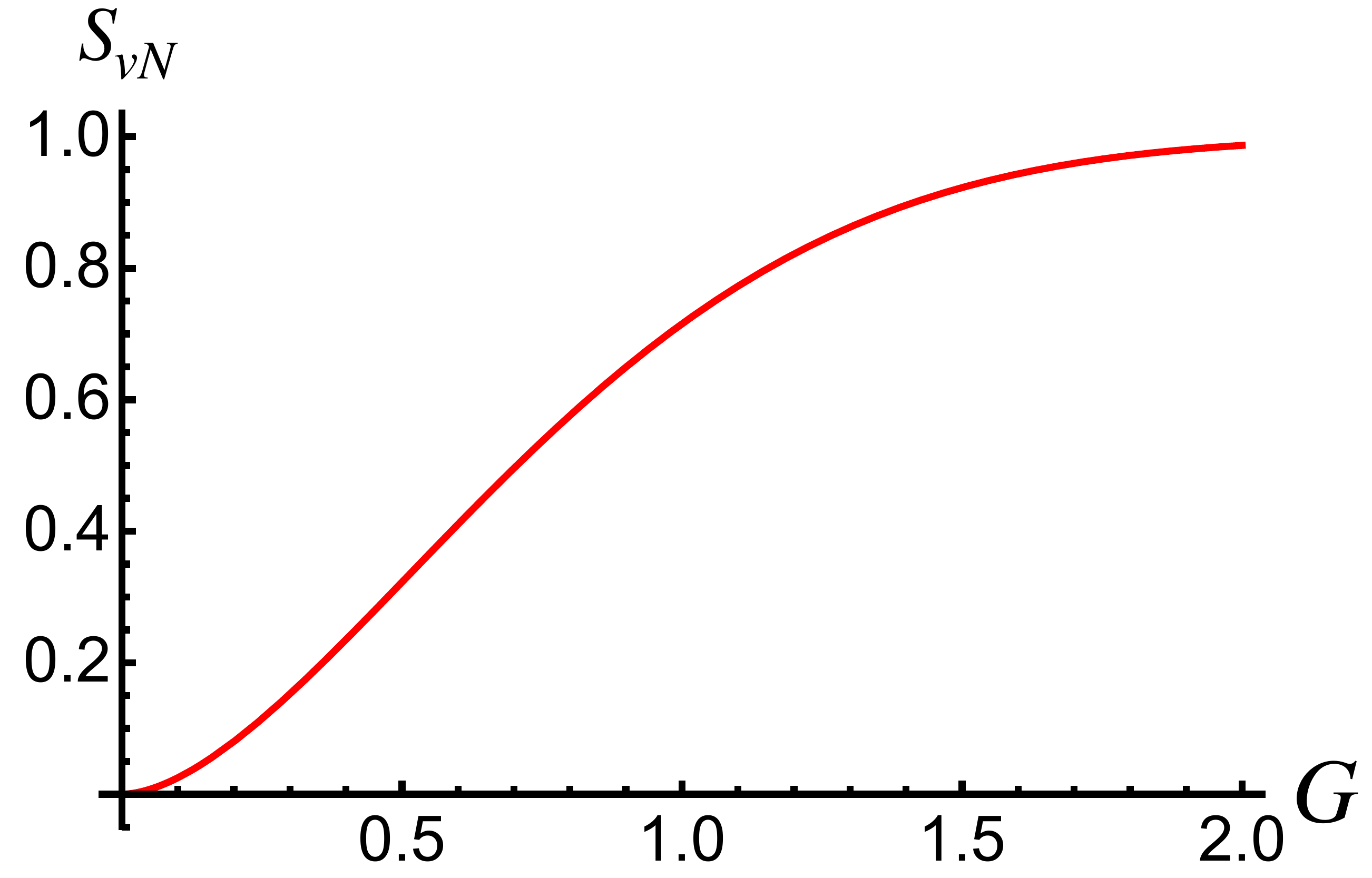}
    \caption{von Neumann entropy $S_{vN}$ of the reduced state of the two-level system resulting from equation~(\ref{evolved}), plotted against the displacement amplitude $G=\eta t$ for $c_{\pm}=1/\sqrt2$.}
    \label{entanglement}
\end{figure}

First, the propagator $\hat{\cal U}_t$ generated by equation~(\ref{hamiltonian}) reads
\begin{equation}
\hat{\cal U}_t=I_{M}\otimes\ket{-1}\bra{-1}_A+\hat {\cal D}_M(-i\eta t)\otimes\ket{+1}\bra{+1}_A
\end{equation}
with $\hat{\cal D}_M(-i\mu)=\exp[-i\mu(\hat b^\dag+\hat b)]$ the displacement operator of amplitude $-i\mu$ along the momentum axis in the single-oscillator phase space. As anticipated, this dynamics realizes a conditional shift of the mechanical state, depending on the state of the two-level system. Using the non-commutative nature of displacement operators of different amplitudes and the Campbell-Baker-Haussdorff formula, we have $\hat{\cal D}_M(-i\mu)\hat{\cal D}_M(\alpha)=\hat{\cal D}_M(\alpha-i\mu) e^{-i\mu{\rm Re}[\alpha]}$, which gives us the evolved state of the atomic-mechanical compound
\begin{eqnarray}
\label{evolved}
\hat{\cal U}_t\ket{\psi_0,\alpha}_{AM}&=c_{-}\ket{-1,\alpha}_{AM}+c_{+}e^{-i\eta t {\rm Re}[\alpha]}\ket{+1,\alpha-i\eta t}_{AM}.
\end{eqnarray}
equation~(\ref{evolved}) is a so-called micro-macro  state that displays, in general, quantum entanglement between the microscopic degrees of freedom of the two-level system and the macroscopic ones of the mechanical mode. The degree of $A$-$M$ entanglement of the evolved state, as measured by the von Neumann entropy of the reduced state of the atom, is independent of the choice of $\alpha$ but only determined by the initial degree of coherence between the atomic states and the degree of distinguishability of $\ket{\alpha}_M$ and $\ket{\alpha-i\eta t}_M$. This ultimately boils down to the value taken by the displacement amplitude. In order to set a benchmark, we take $c_{-}=c_{+}=1/\sqrt 2$, call $G=\eta t$ and get the von Neumann entropy 
\small
\begin{equation}
S_{vN}(G)=-\sum_{s=\pm1}\left(\frac12+ \frac{s}{2} e^{-G^2/2}\right)\log_2\left(\frac12+ \frac{s}{2} e^{-G^2/2}\right).
\end{equation}
\normalsize
Even moderate values of $t$ lead to significant entanglement [cf. figure~\ref{entanglement}]: for instance, we have $S_{vN}(1)\simeq0.72$. 

Such strong quantum correlations between the two-level system and the mechanical one is the key to the emulation of the dynamics in section~\ref{sec:level2a}. While tracing out the two-level system results in the incoherent mixture of coherent states 
\begin{equation}
\hat{\rho}_{M}=|c_{-}|^2|\alpha\rangle\langle\alpha|_M+|c_{+}|^2|\alpha-iG\rangle\langle \alpha-iG|_M,
\end{equation}
which exhibits no quantum feature, the conditional mechanical state achieved by projecting the two-level system onto $\ket{+_x}_A=(\ket{+1}+\ket{-1})_A/\sqrt2$ reads
\begin{equation}
\ket{\psi}_M={N}(c_{-}\ket{\alpha}+c_{+}e^{-iG\alpha}\ket{\alpha-iG})_M
\end{equation}
with ${N}=[c_{-}^2+c_{+}^2+2c_{-} c_{+} \cos(2G\alpha)e^{-G^2/2}]^{-1/2}$. For a sufficiently large value of the displacement $G$, states $\ket{\alpha}_M$ and $\ket{\alpha-iG}_M$ are quasi-orthogonal ($|\langle\alpha\vert\alpha-iG\rangle|^2=e^{-G^2}\simeq0.11$ for $G=1.5$). Therefore, the combination of the joint unitary evolution of the $A$-$M$ compound and projection of the state of the two-level system onto $\ket{+_x}_A$ transfers the coherences that were initially in the state of $A$ to the mechanical mode. This is the key for mimicking the performance at the basis of the test of macrorealism illustrated in section~\ref{sec:level2}.

In what follows, we call 
$\hat{\epsilon}_t$
the dynamical map applied to the initial state of both the mechanical system and the two-level system. The map, resulting from the concatenation of operations illustrated above, gives the time-evolved state of the mechanical system 
\begin{equation}
\label{dynamical}
\hat\rho_{M}(t)={\cal N}\tr_A\left[\hat{\Pi}_{+_x}\hat{\cal U}_t|\psi_0\rangle\langle\psi_0\vert_{A}\otimes\hat\rho_M(0)\hat {\cal U}_t^\dag\right]
\end{equation}
with $\hat\Pi_{+_x}$ the projector onto $\ket{+_x}_A$ and ${\cal N}$ a normalization constant necessary in light of the projective operation. 
This map depends only on the actual time interval in which the system evolves. In this analysis we fixed the value of $G$ meaning that we set a time $t$ so that the actual dynamics of the mirror depends only on the coefficients of the ancillary state.

\section{Macrorealistic test}
\label{sec:level4}

In order to compute the Leggett-Garg inequality we need to find correlation functions for different time intervals. 
Our proposal shares the same interaction mechanism as the one in reference~\cite{asadian}. However, it differs from it significantly, in that our protocol relies on the intuition that the macrorealistic inequality should be violated if the LG function, for the mechanical oscillator, reproduces the dynamic of a spin-1/2 particle whose evolution in time is governed by the $x$-Pauli operator, as shown in section~\ref{sec:level2a}. We accomplish such a goal using the two-level system as an ancilla that assists our scheme at each stage of the macrorealistic test for the oscillator and it is fundamental in obtaining a violation of LG inequality. In reference~\cite{asadian}, a superposition of macroscopically distinguishable states is created with the same effective Hamiltonian as in equation~(\ref{hamiltonian}), but the inequality is tested by performing measurements on the two-level system, whose state carries along a signature of non-classicality for the mechanical oscillator.

We proceed step by step to illustrate the features of our analysis. 

\subsection{Auto-correlation function in time intervals $t_0 - t_1$ and $t_0 - t_3$}
\label{sec:level4a}
The initial state is chosen as a product of a coherent state $\ket{\alpha}_M$ for the mechanical oscillator and the following coherent superposition for the two-level system 
\begin{equation}
|\psi_0\rangle_{A}=(\sin \tau |{+1}\rangle_{A}+\cos \tau |{-1}\rangle_{A});\label{statozero}
\end{equation}
here the coefficients are chosen to be dependent on a parameter $\tau$ which can be manipulated so that the mechanical system will mimic at all times the dynamics of the spin described in section~\ref{sec:level2a}.
With such a choice of initial state of the two-level system, the dynamical map of equation~(\ref{dynamical}) consists of the single Kraus operator
\begin{equation}
\hat{\cal M}_t=\sqrt{{\cal N}_1}\left({\cos\tau}\,I_M+{\sin\tau}\,\hat{\cal D}_M(-iG)\right),
\end{equation}
so that $\hat\rho_M(t_1)=\hat{\cal M}_{t_1}\hat\rho_M(0)\hat{\cal M}^\dag_{t_1}=\ket{\varphi_{t_1}}\bra{\varphi_{t_1}}_M$ with 
\begin{eqnarray}
\label{statotot1}
|\varphi_{t_1}\rangle_{M}
&=\sqrt{\cal N}_{1}\left(\cos\tau|\alpha\rangle_{M}+\sin\tau e^{-iG\alpha}|\alpha-iG\rangle_{M}\right)
\end{eqnarray}
with ${\cal N}_1=[1+\sin(2\tau)\cos(2G\alpha)e^{-G^2/2}]^{-1}$.
It is straightforward to check that this state is similar to the one obtained at time $t_1$ in the case of a spin-1/2 particle. First, using the same notation as in section~\ref{sec:level2a}, the latter reads $
|\phi_{t_1}\rangle= \cos \Delta t|{+1}\rangle - i \sin \Delta t |{-1}\rangle$. Second, in light of the quasi-orthogonality of sufficiently displaced coherent states, we can assume the following logical encoding of a quasi-spin particle into the space spanned by the coherent-state components of equation~(\ref{statotot1}) as follows
\begin{eqnarray}
|\alpha\rangle_M\longrightarrow|+1\rangle_{L},\quad|\alpha-iG\rangle_M\longrightarrow|-1\rangle_{L}.
\end{eqnarray}
Here, the subscript $L$ stands for the logical two-level system that we have invoked. Notice that similar encodings represent the building blocks of coherent-state quantum computing~\cite{Kok2007}. Finally, as we will see in the remainder of our analysis, the presence of the extra relative phase between the logical states is inessential for the success of our test. 

With such encoding, the observable of choice for the construction of the two-time auto-correlation functions entering the LG function akin to equation~(\ref{LGI}) would be  given by $\ket{+1}\bra{+1}_L-\ket{-1}\bra{-1}_L$. This implies the ability to discriminate between $\ket{\alpha}_M$ and $\ket{\alpha-iG}_M$, which can be done as discussed in reference~\cite{Jeong2002} [cf. section~\ref{resu}]

\begin{figure}[t]
\centering{\bf (a)} \hskip5cm{\bf (b)}\\
\centering
    \includegraphics[width=0.4\columnwidth]{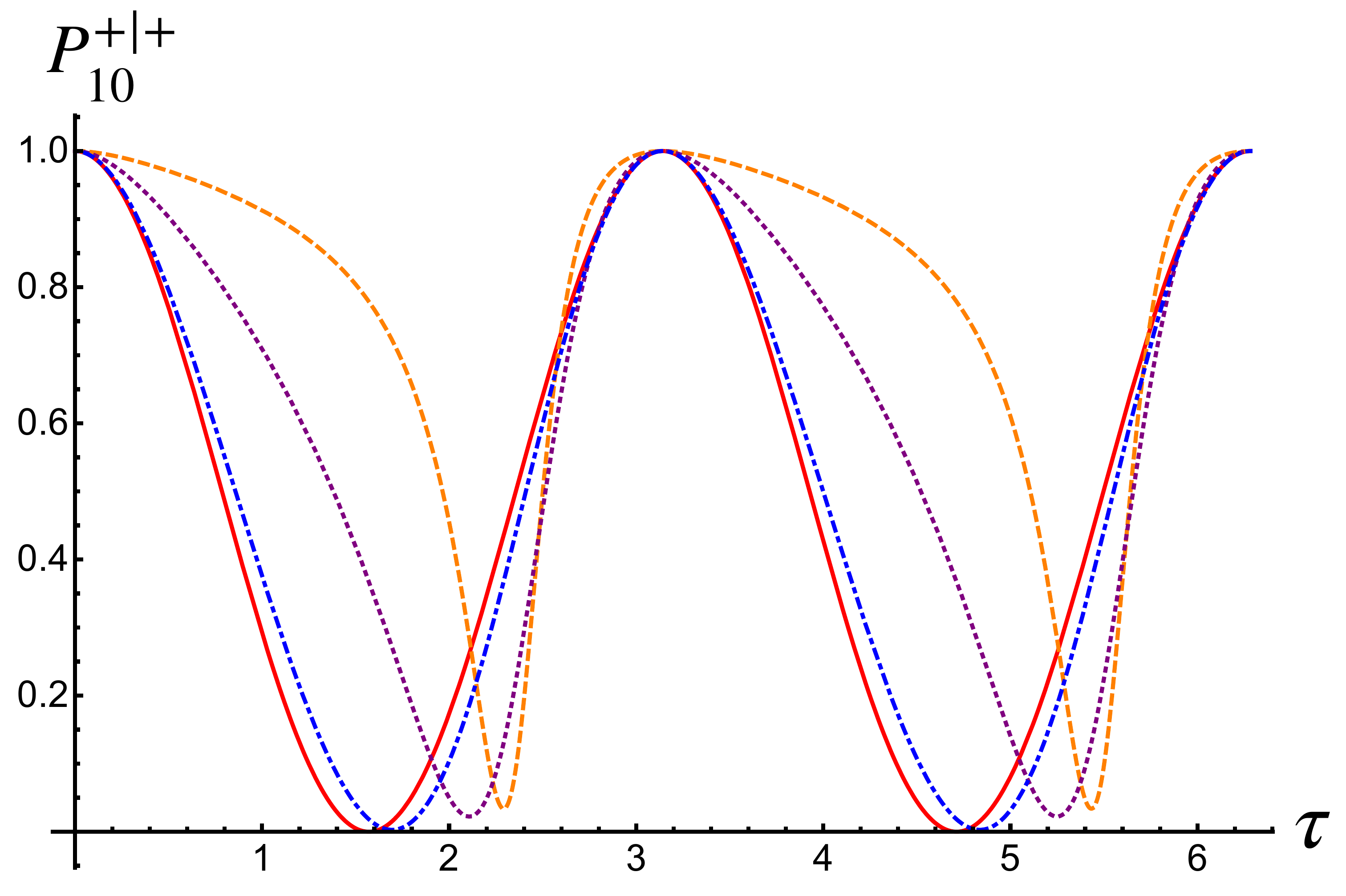}\includegraphics[width=0.371\columnwidth]{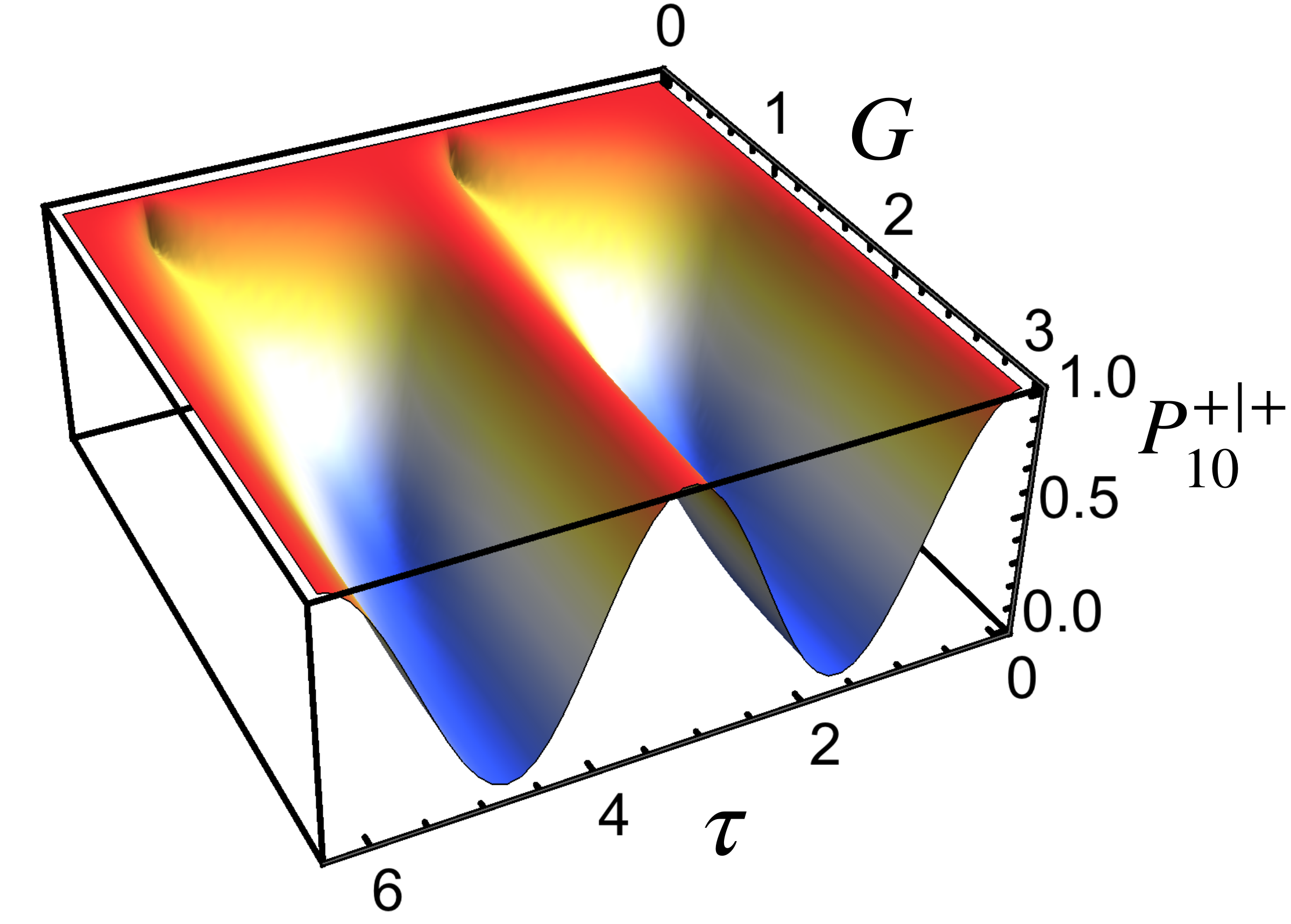}
    \caption{{\bf (a)}: Conditional probability $P_{10}^{+|+}$ plotted against $\tau$ for $\alpha=0.1$ and $G=0.5$ (dashed orange line), $G=1.0$ (dotted purple line), and $G=2$ (dot-dashed blue line). The solid red curve shows the behavior of the analogous conditional probability for the case of a spin-1/2 particle discussed in section~\ref{sec:level2a}. {\bf (b)}: We study the convergence of $P_{10}^{+|+}$ towards the benchmark $(\cos\tau)^2$ form of such conditional probability as $G$ grows for $\alpha=1$.}
    \label{t0t1}
\end{figure}

We can now compute the probabilities to build the two-time  correlation functions. We have
\begin{equation}
P_{0}^+=1,\quad P_{0}^{-}=1 -P_{0}^+=0.\label{prob0}
\end{equation}
The conditional probability to obtain outcome $+1_L$ at time $t_{1}$ provided we obtained the same outcome at $t_{0}$ is
\begin{eqnarray}
\label{probc01_1}
P_{10}^{+|+}&={}_M\langle\alpha|\hat{\epsilon}_{t_1}(|\alpha\rangle\langle\alpha|_M)|\alpha\rangle_M\nonumber\\
&=|{}_M\langle\varphi_{t_1}\vert\alpha\rangle_M|^2\\
&= {\cal N}_{1}\left |e^{-\frac{G^{2}}{2}-2 i G \alpha}\sin\tau+\cos\tau\right|^{2}.\nonumber
\end{eqnarray}
This expression shows that a sufficiently large distinguishability of the coherent-state components of $\ket{\varphi_{t_1}}$ (i.e. a faithful encoding of a two-level system) gives $P_{10}^{+|+}\to\cos^2\tau$, thus reproducing the result valid in the spin-1/2 case. Notice that the value of $G$ at which this occurs depends on the amplitude $\alpha$ of the initial coherent state of system $M$. This is shown in figure~\ref{t0t1}, where we address the features of $P_{10}^{+|+}$ against both $\tau$ and $G$. 

Needless to say, the conditional probability to find $-1_L$ at $t_{1}$ once $+1_L$ has been achieved at $t_{0}$ is obtained by conservation of total probability as
$P_{10}^{-|+}=1-P_{10}^{+|+}$.
The two-time correlation function from the initial time $t_{0}$ and the time $t_{1}$ is thus given by the overall expression
\small
\begin{eqnarray}
\label{Ct0t1}
C_{01}=&-1+2\frac{e^{-\frac{G^2}{2}} [\sin(2\tau)\cos (2 \alpha  G)+\sin ^2\tau]+\cos ^2\tau}{1+e^{-\frac{G^2}{2}}\cos(2\alpha G)\sin(2\tau)}.
\end{eqnarray}

\normalsize
For $G\gg{1}$, the correlation function in equation~(\ref{Ct0t1}) tends to $\cos(2\tau)$, thus recovering that of the simple spin-1/2 model. 
Notice that the evaluation of the two-time auto-correlation function for the time period $t_0-t_3$ proceeds along lines that are very similar to those followed in order to evaluate $C_{01}$. In fact, it is enough to change $\tau\to3\tau$ and $G\to3G$ in equation~(\ref{Ct0t1}) to get $C_{03}$.

\subsection{Auto-correlation function in time intervals $t_1 - t_2$ and $t_2 - t_3$}
\label{sec:level4b}
For the next time interval, we need to reset the initial state of the two-level system, while system $M$ is in the state given in equation~(\ref{statotot1}).
The state at the instant of time $t_{1}$ is thus
\begin{eqnarray}
    {\cal N}_{1}\left(\cos\tau|\alpha\rangle+\sin\tau e^{-iG\alpha}|\alpha-iG\rangle\right)_M\otimes|\psi_1\rangle_A
\label{statoSMuno}
\end{eqnarray}
with $|\psi_1\rangle_A=A_{\tau}|{+1}\rangle_{A}+B_{\tau}|{-1}\rangle_{A}$. The coefficients $A_{\tau}$ and $B_{\tau}$ depend on the value of $\tau$ and will be determined by imposing the constraint that the evolved state of $M$ at  $t_{2}$ must have the same shape of the one for the spin-1/2 particle, that is $|\phi_{t_2}\rangle=\cos (2\Delta t)|{+1}\rangle - i \sin (2\Delta t) |{-1}\rangle$.

Proceeding as done in section~\ref{sec:level4a} we get to the reduced state of the mechanical system $M$ at time $t_2$
\begin{eqnarray}
\label{statodue}
|\varphi_{t_2}\rangle_{M}&={\cal N}_1{\cal N}_2\big[B_{\tau}\cos\tau |\alpha\rangle_M+A_{\tau}e^{-2iG\alpha}\sin\tau |\alpha-2iG\rangle_M \nonumber\\
&+\left(A_{\tau}\cos\tau+B_{\tau}\sin\tau\right) e^{-iG\alpha}|\alpha-iG\rangle_M\big].
\end{eqnarray}
The explicit form of ${\cal N}_2$, $A_\tau$ and $B_\tau$ is straightforward to obtain by ensuring the normalization of $\ket{\psi_1}_A$ and $\ket{\varphi_{t_2}}_M$ and imposing the condition 
\begin{equation}
B_\tau{\cal N}_1{\cal N}_2\cos\tau=\cos(2\tau),
\end{equation}
which would render equation~(\ref{statodue}) analogous to the corresponding one in the spin-1/2 case. Such expressions are too cumbersome to be reported here. 

The presence of $\ket{\alpha-2iG}$ in equation~(\ref{statodue}) paves the way to considerations on the logical encoding chosen for our scheme, which we now arrange in such a way that the 
following positive valued operator measurement (POVM) is considered 
\begin{equation}
\hat \Pi_+=\ket{\alpha}\bra{\alpha},\qquad\hat\Pi_{-}=I-\ket{\alpha}\bra{\alpha}
\end{equation}
with $\hat\Pi_{+}$ the projector onto the logical $\ket{+1}_L$ state.

{The two-time autocorrelation function between $t_2$ and $t_3$  can be computed in a way analogous to what has been done in the previous time interval, making use of the following conditional probabilities and their complements
\begin{equation}
P_{21}^{+|+}=\bigr|\;{\cal N}_1{\cal N}_{2}\bigl(\;e^{-\frac{G^{2}}{2}-2iG\alpha}\sin\tau+\cos\tau\;\bigr)^{2}\;\bigr|^{2},\label{probc12_1}
\end{equation}
\begin{eqnarray}
P_{21}^{+|-}&=|{\cal N}_1{\cal N}_{3}|^2\left |(\sin\tau)^2 e^{-2iG\alpha}(e^{-2G^{2}}{-}e^{-G^{2}})
+\frac{\sin(2\tau)}{2} (1-e^{-\frac{G^{2}}{2}})\right|^{2}.\label{probc12_3}
\end{eqnarray}}

{Calculations for the last time interval, $t_2$-$t_3$ have been done following the same procedure illustrated above. The corresponding expressions are, however, not informative enough to be reported here, and we thus omit them.}

\subsection{Results of the test}
\label{resu}

The Leggett-Garg function $K$ can finally be computed to test the inequality
\begin{equation}
K=|C_{01}+C_{12}+C_{23}-C_{03}|\leq2
\end{equation}
against the coupling parameter $G$. Larger values of $G$ mean that the encoded basis of coherent states is more similar to an orthogonal basis, thus making function $K$ similar to the one for the spin-$1/2$ particle. The main difference between the two cases rests in the existence of a natural dichotomic observable upon which to perform measurements in the latter case, which is not the case for a harmonic oscillator.

\begin{figure}[t]
\centering
\includegraphics[width=0.65\columnwidth]{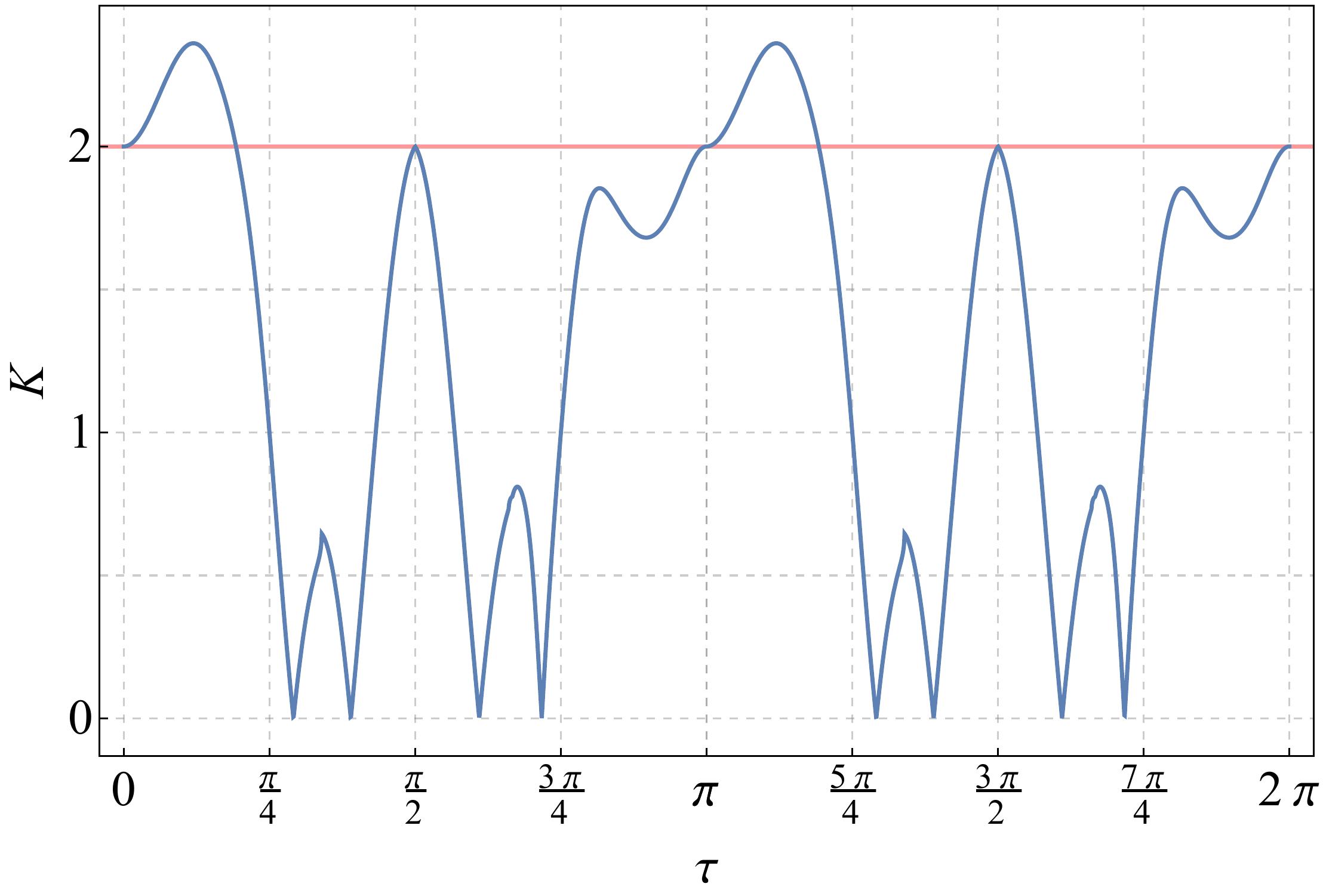}\caption{Plot of the Leggett-Garg function for the oscillator when the integrated coupling constant is $G=5$. The value of the parameter $\alpha=1$ has been chosen from these calculations. The maximum value $K=2.36$ is reached for $\tau=0.37$. }\label{LGplot2}
\end{figure}

Notwithstanding such fundamental difference, the LG function for the harmonic oscillator in figure~\ref{LGplot2} closely resembles the spin-1/2 case even for moderate values of $G$. Function $K$ shows a periodicity of $\pi$ in each plot, while in the previous case the period was ${\pi}/{2}$; this is due to the difference in the correlation functions involved in its definition. 
The violation of the macrorealistic inequality at short evolution times shown in figure~\ref{LGplot2} signals the non-classical character of the statistics sampled in order to construct the conditional probabilities entering $K$.

It is worth mentioning that the correlation function that we need to reconstruct the LG function could be inferred by performing measurements on a probing cavity field. A possibility is to use a double-optomechanical cavity with the mechanical oscillator in a membrane-in-the-middle configuration, as suggested in reference~\cite{vitali}. A second possibility is to use a single cavity, as in figure 2, and a second mode with a polarization that is orthogonal to the one of the field used to mediate the interaction between the two-level ancilla and the mechanical oscillator in our scheme. By arranging suitable conditions of adiabatic following between mechanical oscillator and probing light field, as illustrated in reference~\cite{vitali}, one can map the temporal behaviour of the relevant operators of the mechanical oscillator onto that of the probing field. Noticeably, the scheme works under the assumption of large damping rate for the probing field mode. The measurement scheme would thus be robust against the cavity dissipation rate. Measurements of temporal correlations on the latter, and the calibration of the noise affecting the cavity, would thus result in the possibility to infer the violation of the LG inequality.

The last ingredient of the measurement scheme consists of the distinction between coherent states, which enables the logical encoding invoked in our work. Such distinction can be done using the probing optical field upon which we would write the state of the mechanical oscillator and interference of its coherent state to a reference one at a 50:50 beam splitter, following the protocol illustrated in reference~\cite{Jeong2002}. Schemes for the direct tomography of the mechanical state have also been put forward~\cite{Vanner}.

\subsection{Open system dynamics}
\label{sec:level3c}

\begin{figure}[t]
\centering
{\bf (a)}\hskip4.5cm{\bf (b)}\\
\centering
\includegraphics[width=0.345\columnwidth]{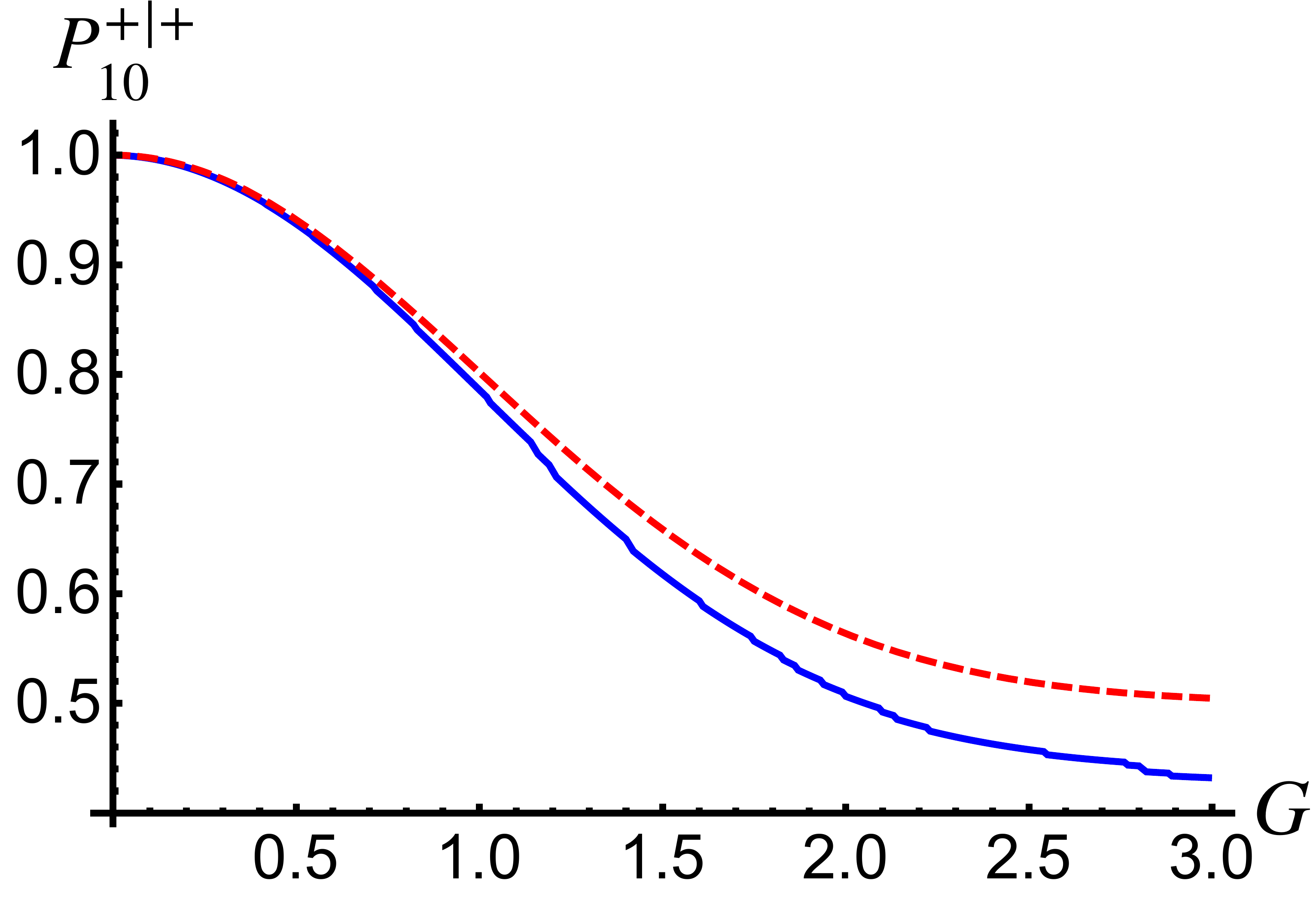}\includegraphics[width=0.37\columnwidth]{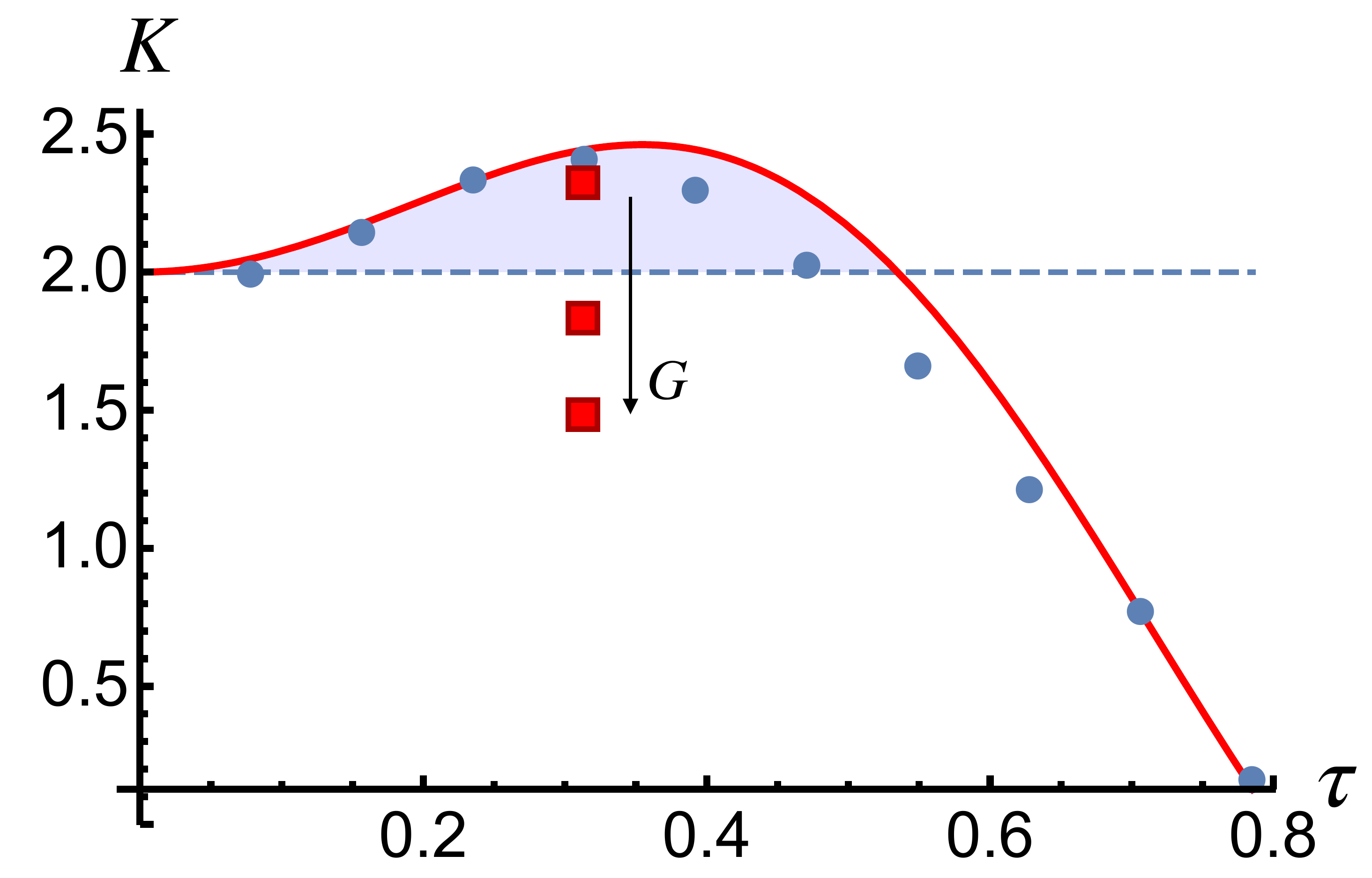}
\caption{{\bf (a)} We show the conditional probability $P^{+|+}_{10}$ against the values of $G$ for a unitary evolution (red dashed line) and a conservative value of $\kappa/\eta=10^{-3}$ (blue solid line). {\bf (b)} Leggett-Garg function $K$ against $\tau$ for the ideal unitary dynamics (red solid line) and $\kappa/\eta=10^{-3}$ with $G=2$ (blue dots). The red squares show the value taken by $K$  at $\tau=\pi/10$ for $\kappa/\eta=10^{-3}$ and $G=2.5,3,3.5$ (growing as shown by the arrow). The dashed horizontal line shows the classical macrorealistic bound and the shaded region highlights the range of values of $\tau$ where the LG inequality is violated. Assuming that $\eta$ lies between 1 and 100 Hz \cite{Vacanti2013}, the value of $\kappa/\eta$ suggests a quality factor $Q=\omega/\kappa$ in the range $10^7 - 10^9$ which is experimentally feasible \cite{Aspelmeyer}. }
\label{open}
\end{figure}

In order to account for the potentially detrimental effects due to the unavoidable environmental action, we have conducted an investigation including mechanical dissipation. This has been done by assuming Markovian damping in a cold environment at a rate $\kappa$, resulting in 
 the master equation
\begin{equation}
\label{masterEq}
    \frac{d \hat{\rho}}{dt} = -i [\hat{H}_{eff},\hat{\rho}] + \frac{\kappa}{2} (2 \hat{b} \hat{\rho}\hat{b}^\dagger - \hat{b}^\dagger \hat{b} \hat{\rho} - \hat{\rho}\hat{b}^\dagger \hat{b}),
\end{equation}
where 
$\hat{\rho}$ is the state of the system. 
We solved equation~(\ref{masterEq}) using the quantum unravelling approach~\cite{Dalibard1992}. This involves dividing the time interval into very small time steps of length $\delta t$. A random number between 0 and 1 is generated at the start of each time step and we calculate the value of the function
\begin{equation}
\delta p = \delta t \kappa _{AM} \bra{\varphi}  \hat{b}^\dagger \hat{b} \ket{\varphi}_{AM},
\end{equation}
where $\ket{\varphi}$ is the state of the atom and mechanical oscillator at this stage of the evolution. The value of this function will always be much smaller than 1 because it depends on the small time step $\delta t$. If the random number is greater than $\delta p$, which it will be in most cases, then the system evolves according to a modified Hamiltonian, which for this system reads 
\begin{equation}
    \hat{H}_\mathrm{QU} = \hat{H}_{eff} + \frac{i \kappa}{2} \hat{b}^\dagger \hat{b}.
\end{equation}
Since the time steps are very small, we can evolve the system according to the time evolution operator
\begin{equation}
    \hat{U}(\delta t) = e^{-i \hat{H}_{\rm QU} \delta t} \sim \hat{I} - i \hat{H}_{\rm QU} \delta t.
\end{equation}
If the random number is less than $\delta p$, then a quantum jump occurs. Here, the quantum jump operator is $\sqrt{\kappa} \hat{b}$, so when this acts on the state of the system the mechanical oscillator effectively loses one excitation. After each time step the state must be renormalized. We evolved the system in this way until a certain final time and obtained a trajectory, or a possible outcome of this process. We repeated this many times and collected many trajectories before averaging over them to obtain the solution of equation~(\ref{masterEq}). Once we obtained the state of the system at time $t$, we projected the state of the two-level system onto $\ket{+_x}_A$ and found the reduced state of the mechanical oscillator as in section~\ref{sec:level3b}.

The results of this approach are shown in figure~\ref{open} {\bf (a)} and {\bf (b)}, where we study the decay of the probability $P^{+|+}_{10}$ as a paradigmatic instance of the effects of the mechanical damping. Large values of $G$, despite ideally making the mechanical state components more distinguishable, imply larger evolution times of the mechanical oscillator and thus more chances of environmental action. This results in the spoiling of the damping-affected $P^{+|+}_{10}$ with respect to the ideally closed case scenario. This has an effect over the values taken by the LG function $K$, as studied in figure~\ref{open} {\bf (b)}: the range of values of $\tau$ within which we would observe a violation of the LG inequality is shrunk (while the amplitude of violation is slightly reduced). Moreover, at sufficiently large values of $G$ we lose the violation of the macrorealistic inequality altogether.

\section{Conclusions}
\label{sec:level5}

We have proposed a method to violate a macrorealistic LG-like inequality in a hybrid optomechanical setting. By making use of the control allowed by the engineered interaction between a mechanical oscillator and an ancillary two-level system, we have shown a protocol able to mimic closely the features of the conditional probabilities entering the LG function of a spin-1/2 particle. The success of our scheme relies on the ability of the ancilla-oscillator interaction to generate coherent superpositions of distinguishable mechanical states, whose features are adjusted to achieve a significant violation of the classical macrorealistic bound. We have assessed the robustness of the violation to the effects of environmental damping, showing that it strongly depends on suitable arrangements of the ancilla-oscillator interaction time. The scheme is robust against a residual thermal character of the initial preparation of the mechanical oscillator as far as the width of the Wigner function of the initial thermal state in phase space is smaller than the value of $G$. By exploiting a logical encoding reminiscent of the one used in the context of coherent state-based quantum information processing, our proposal contributes to the endeavours for the design of viable routes toward the observation of non-classical effect at the mesoscopic and macroscopic scale. 

\ack
MM, AB, and MP gratefully acknowledge support by the H2020 Collaborative Project TEQ (Grant Agreement 766900). HM and MP are supported by the SFI-DfE Investigator Programme through project QuNaNet (grant 15/IA/2864). MP also acknowledges support from the Leverhulme Trust through the Research Project Grant UltraQuTe (grant nr.~RGP-2018-266) and the Royal Society Wolfson Fellowship scheme through project ExTraQCT (RSWF\textbackslash R3\textbackslash183013).


\end{document}